\documentclass[a4paper,aps,prl,twocolumn,showpacs]{revtex4-1}
\setcitestyle{super}
\usepackage{graphicx}
\usepackage[utf8]{inputenc}
\usepackage{amsmath,amssymb,times}
\usepackage{subfigure,color,url}
\usepackage{lineno,empheq}

\usepackage[pdfstartview=FitV,bookmarks=true,
            linktocpage=true,colorlinks=true,
            urlcolor=blue,linkcolor=blue,
            citecolor=blue]{hyperref}

\graphicspath{{Images/}}
\begin{document}

%\preprint{\textit{published in Wave Motion}}
\title{Temporal behavior of laser induced elastic plate resonances}
\author{J\'{e}r\^{o}me Laurent} \email{jerome.laurent@espci.fr}
\author{Daniel Royer}           \email{daniel.royer@espci.fr}
\author{Claire Prada}           \email{claire.prada@espci.fr}
\affiliation{Institut Langevin, ESPCI ParisTech, CNRS,\\ 1 rue Jussieu, 75238 Paris Cedex 05, France}
%\date{\today}

\begin{abstract}
This paper investigates the dependence on Poisson's ratio of local plate resonances in low attenuating materials. In our experiments, these resonances are generated by a pulse laser source and detected with a heterodyne interferometer measuring surface displacement normal to the plate. The laser impact induces a set of resonances that are dominated by Zero Group Velocity (ZGV) Lamb modes. For some Poisson's ratio, thickness-shear resonances are also detected. These experiments confirm that the temporal decay of ZGV modes follows a $t^{-0.5}$ law and show that the temporal decay of the thickness resonances is much faster. Similar decays are obtained by numerical simulations achieved with a finite difference code. A simple model is proposed to describe the thickness resonances. It predicts that a thickness mode decays as $t^{-1.5}$ for large times and that the resonance amplitude is proportional to $D^{-1.5}$ where $D$ is the curvature of the dispersion curve $\omega(k)$ at $k=0$. This curvature depends on the order of the mode and on the Poisson's ratio, and it explains why some thickness resonances are well detected while others are not. 
\end{abstract}


% \linenumbers

\maketitle

% ================
%   Introduction
% ================

\section{Introduction}

Small amplitude vibrations of an elastic plate are governed by the three-dimensional equations of the linear theory of elasticity. For an infinite homogeneous plate, the frequency $f = \omega/2\pi$ depends on the wavelength $\lambda=2\pi/k$ in the plane of the plate. When the faces of the plate are free of traction, no energy leakage occurs, then for any real $k$, the secular equations established by Rayleigh~\cite{Rayleigh89} yield an infinite number of real roots in $\omega$. The dispersion curves of these symmetric $(S_n)$ and anti-symmetric $(A_n)$ propagating modes, guided by the plate, are represented by a set of branches in the $(\omega, k)$-plane.\citep{Lamb17,Mindlin56,Mindlin06} The complete Lamb mode spectrum depends on material parameters, either expressed by the longitudinal to transverse wave velocity ratio $V_L/V_T$ or by the Poisson's ratio $\nu$.\cite{Royer09} Dispersion curves $\omega(k)$ of high order modes start from the $k = 0$ axis at a finite ordinate $\omega_c$. At these cut-off frequencies $f_c = \omega_c/2\pi$ multiple reflections of longitudinal or shear waves between the top and bottom faces of the plate, give rise to thickness-shear resonances (modes $S_{2n}$ or $A_{2m+1}$) or to thickness-stretch resonances (modes $S_{2m+1}$ or $A_{2n}$) at infinite wavelength.\\

\begin{figure*}[!ht]
\centering
\begin{minipage}[c]{0.77\textwidth}
\subfigure{\includegraphics[width=0.9\textwidth]{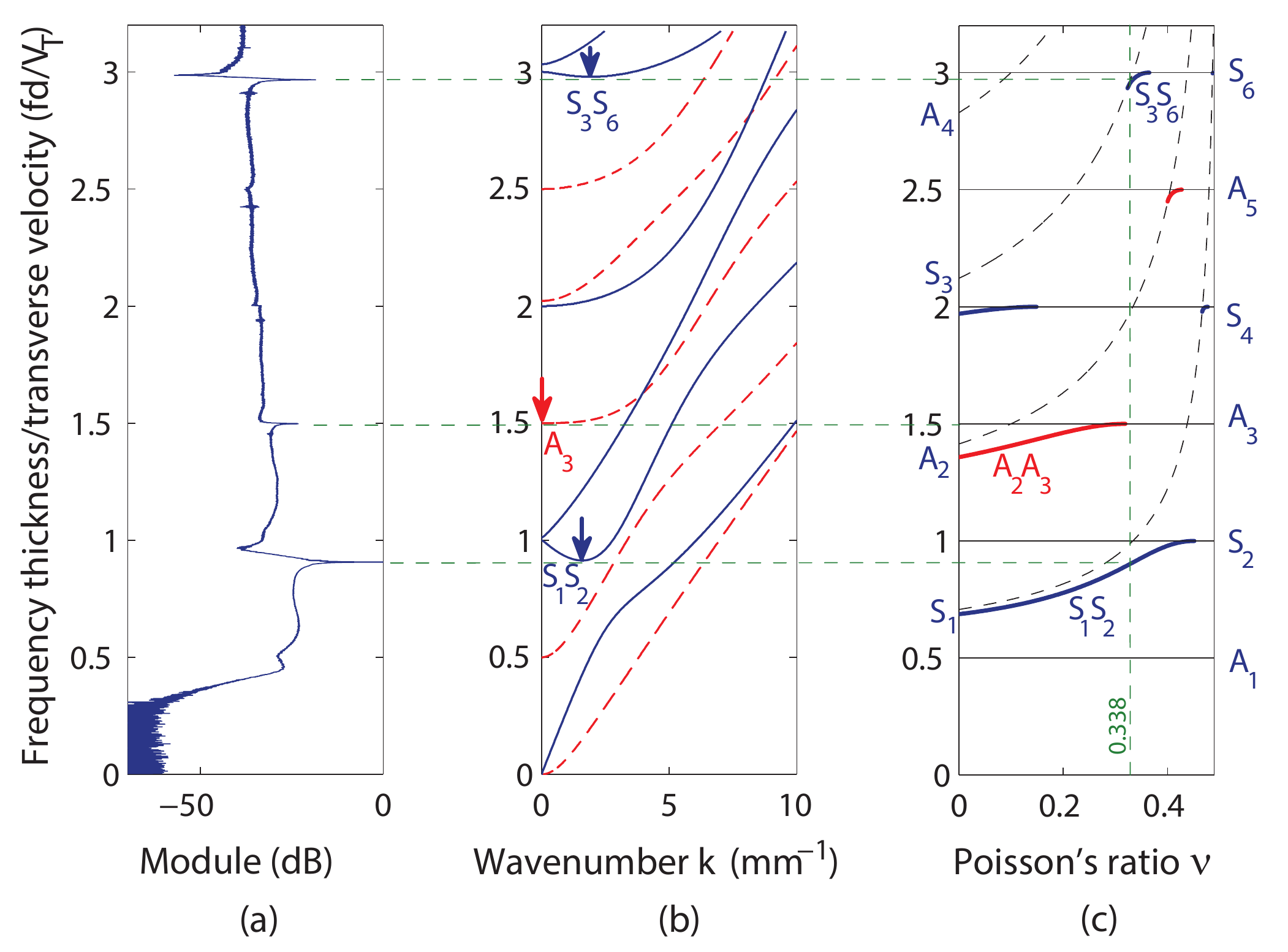}\label{ZGVthSpectreDispersion1}}
\subfigure{\label{ZGVthSpectreDispersion2}}
\subfigure{\label{ZGVthSpectreDispersion3}}
\end{minipage}\hfill
\begin{minipage}[c]{0.23\textwidth}
\caption{(a) Spectrum of the normal vibration of a Duralumin plate of thickness $d=1$~mm
generated and detected by laser. Vertical scale: $fd/V_T$. (b) Normalized dispersion curves for a Duralumin plate with bulk wave velocities $V_L=6370$~m/s and $V_T=3150$~m/s ($\nu = 0.338$). (c) Dimensionless cutoff frequencies and minimum frequencies of Lamb modes versus Poisson's ratio (horizontal line: thickness-shear modes, dashed line: thickness-stretch modes, and thick line: ZGV modes).}
\label{ZGVthSpectreDispersion}
\end{minipage}
\end{figure*}

In many applications, the plate undergoes a local and impulsive excitation. The spectra of the transient waves in an elastic plate has been analyzed by Weaver and Pao.\cite{Weaver82a} and Santosa~\citep{Santosa89} When surface stresses are induced by a laser source, the coupling of the thermo-elastic source with elastic waves is a complex problem.\cite{Rose84} The theory and simulation of laser generated waves propagating in plates has been the object of several studies.\citep{Spicer90,Cheng01} In a recent paper, Laguerre and Tresseyde~\cite{Treyssede13} proposed a method to calculate the excitability of both propagating and non propagating modes. In practice, the energy deposited on the plate by the source of finite dimensions rapidly flows out of the source area except for non propagative modes. Zero group velocity (ZGV) modes were observed experimentally with various techniques: air-coupled transducers,\cite{Holland03} impact echo method,\cite{Gibson05} laser ultrasonics.\citep{Prada05a,Clorennec06} These experiments demonstrate that the local vibration spectrum of a free plate is dominated by the resonance at the minimum frequency of the $S_1$ Lamb mode. This frequency is slightly lower than the fundamental thickness frequency $f_c = V_L/2d $ and corresponds to the junction of $S_1$ and $S_{2b}$ branches, where $b$ stands for backward wave.\citep{Meitzler65} This is why we chose the notation $S_1S_2$-ZGV resonance.\citep{Ces11} In fact, except for the first three ($S_0$, $A_0$ and $A_1$) Lamb modes, all higher order modes exhibit a minimum frequency for some Poisson's ratio.\citep{Shuvalov08,Prada08b} Frequency minima always occur below the cut-off and correspond to wavelengths of the order of the plate thickness. Because of there finite wavelength, ZGV modes dominate the frequency spectrum of the normal surface displacement after a local impact. However, it was observed that some thickness resonances (infinite wavelength) are also detected when the signal to noise ratio is sufficiently high. For example, Fig.~\ref{ZGVthSpectreDispersion1} displays the resonance spectrum measured at the source point (source spot 2.5 mm), on a 1 mm-thick Duralumin plate. The three observed resonances can be identified from the dispersion curves shown in Fig.~\ref{ZGVthSpectreDispersion2}. The first and the third ones correspond to the $S_1S_2$ and the $S_3S_6$-ZGV modes while the second one is associated to the $A_3$ thickness-shear mode. Fig.~\ref{ZGVthSpectreDispersion3} presents the dimensionless cutoff frequencies and minimum frequencies of Lamb modes versus the Poisson's ratio: the horizontal lines correspond to thickness-shear modes, the dashed lines to thickness-stretch modes and the thick lines to ZGV modes. For $\nu=0.338$ (vertical line), only two ZGV resonances exist at normalized frequencies below 3.2. It appears that the $A_3$ thickness-shear resonance, although 20~dB below the $S_1S_2$-ZGV resonance, is clearly detected, while for example the $A_1$ or the $A_5$ thickness-shear resonances are not observed.\\

The objective of this paper is to explain these observations using theory, simulation and experiments. It is organized as follows: In Sec.~\ref{sec:1}, a simple model is proposed to estimate the temporal decay of ZGV and thickness-shear resonances. The relative amplitudes of thickness resonances, excited and detected by laser techniques, are predicted through simple approximations. Then, Sec.~\ref{sec:2} presents experimental results obtained on Duralumin and fused silica plates. The temporal decay of the different resonances are observed and compared to those obtained with a finite difference code.

% ======================
%   Theoretical model
% ======================
\section{Analysis of the resonance temporal decay}\label{sec:1}

In order to compare the ZGV and thickness resonance behavior, we propose a simple approach only valid for low attenuation materials. The temporal decay of the $S_1S_2$-ZGV Lamb mode resonance was studied in Prada et al.\cite{Prada08a} In this paper [Eq.~(3)], the normal surface displacement associated to a given Lamb mode was expressed as
\begin{equation}
	u(r,t)=\frac{1}{2\pi} \int_0^{+\infty}{ C_{th}(k)Q(\omega)B(k)J_0(kr)e^{i\omega t} k \,dk},
	\label{integral}
\end{equation}
where $C_{th}(k)$ is the coefficient of thermo-elastic conversion into the normal displacement of the mode, $Q(\omega)$ is the spectral content of the laser pulse, and $B(k)$ is the spatial Fourier transform of the distribution of energy $b(r)$ deposited on the surface. For a Gaussian source beam of radius $R$ at $1/e$ 
\begin{equation}
	b(r) = \frac{E}{\pi R^2}\exp\left(-\frac{r^2}{R^2}\right),
	\label{TermSource}
\end{equation}
where $E$ is the deposited energy. The spatial 2D-Fourier transform is then 
\begin{equation*}
	B(k)=B(0)\exp(-sk^2),
\end{equation*}
with $s=R^2/4$ and $B(0)=E/\sqrt{\pi R^2}$. This integral can be calculated for large times $t$ using a Taylor expansion of the dispersion relation $\omega(k)$ in the vicinity of a resonance frequency. As the normal surface displacement vanishes for thickness-shear resonances and not for ZGV resonances, the solution is derived in a different manner for each type of resonances. 

% -----------------------------------------------------------------
\subsection{Normal displacement for a ZGV resonance}\label{sec:11}
% -----------------------------------------------------------------

\begin{figure}[!ht]
\centering
\includegraphics[width=0.8\columnwidth]{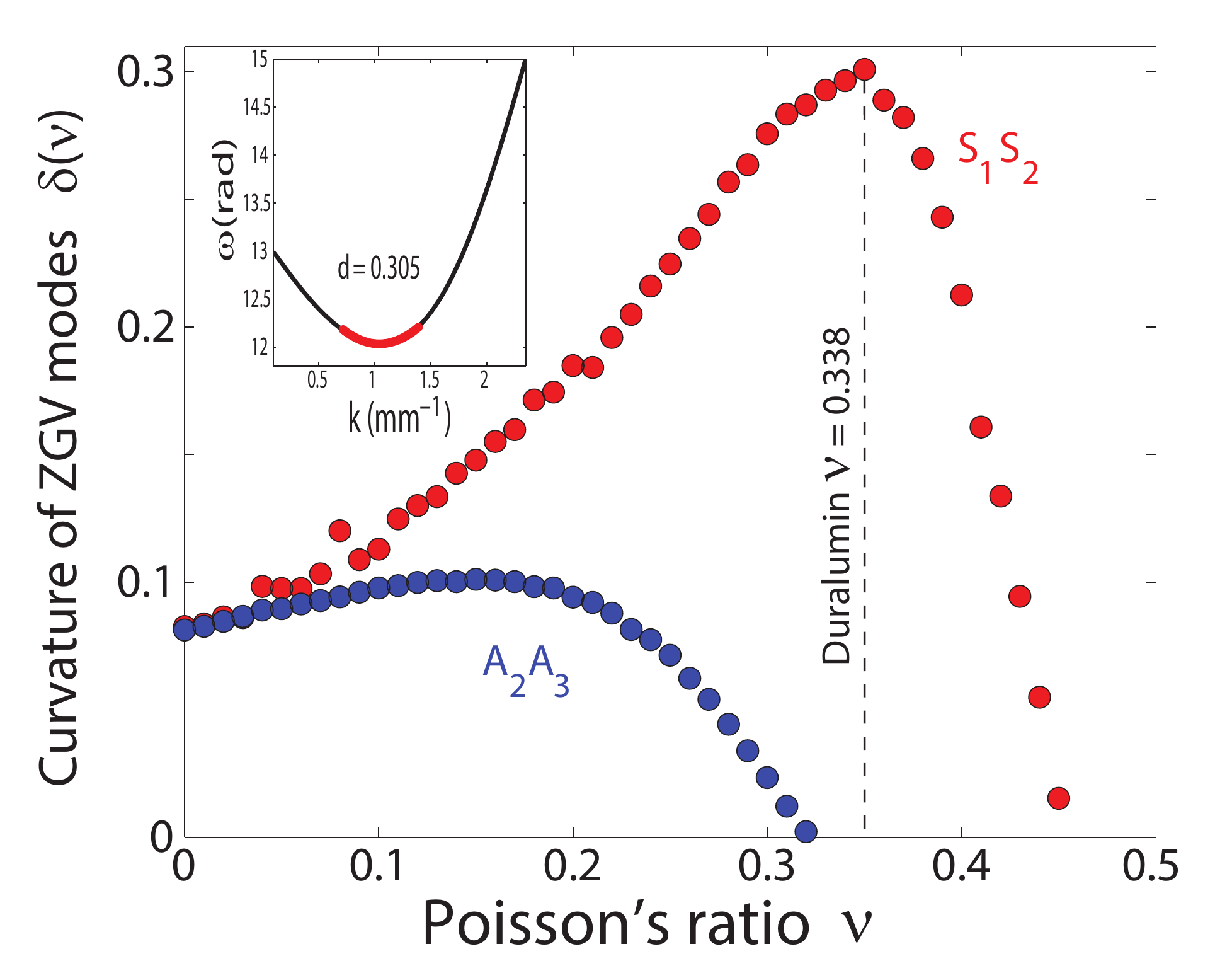}
\caption{Curvature of $S_1S_2$ and $A_2A_3$-ZGV modes as a function of the Poisson's ratio. Inset, dispersion curve of $S_1S_2$ mode in a Duralumin plate. Fit (red line) by Eq.~\eqref{wk:fit}.}
\label{curvature}
\end{figure}

The result established in~\cite{Prada08a} is recalled. In the vicinity of a zero group velocity point $(k_0,\omega_0)$, the second order Taylor expansion of the dispersion curve is written
\begin{equation}
	\omega(k) = \omega_0 + D(k-k_0)^2 + O[(k-k_0)^3],
	\label{wk:fit}
\end{equation}
where the curvature $D$ at ZGV point depends on the mode order and on the Poisson's ratio. Then the integral of Eq.~\eqref{integral} can be approximated by the stationary phase method as
\begin{equation}
	u(r,t) = \frac{C_{th}(k_0)}{\sqrt{4\pi Dt}} Q(\omega_0) B(k_0) J_0(k_0r) k_0 e^{i\left(\omega_0 t+\frac{\pi}{4}\right)} \sim (Dt)^{-0.5}
	\label{ZGVdecay}
\end{equation}
It appears that the temporal decay of the resonance follows a $t^{-0.5}$ law. Furthermore, the resonance amplitude is proportional to the inverse square root of the curvature of the dispersion curve $D$. The coefficient $D$ normalized to the product of the transverse wave velocity by the plate thickness is a dimensionless coefficient $\delta(\nu)=D/V_Td$ which only depends on the Poisson's ratio $\nu$. This parameter, numerically calculated for the first two ZGV modes $S_1S_2$ and $A_2A_3$, is displayed in Fig.~\ref{curvature}. It is interesting to notice that for the $S_1S_2$ mode, $\delta(\nu)$ is maximum for a Duralumin plate ($\nu=0.338$) and approximately equal to 0.305.
%

% -----------------------------------------------------------------------------
\subsection{Normal displacement for a thickness-shear resonance}\label{sec:12}
% -----------------------------------------------------------------------------

In the vicinity of a cut-off frequency $f_c=\omega_c/2\pi$, the slope of the dispersion curve generally vanishes. As shown by Mindlin,\cite{Mindlin06} this is not true when there is a coincidence between a thickness-shear and a thickness-stretch resonance of the same symmetry. The coincidence occurs for symmetrical modes $S_{2m+1}$ and $S_{2n}$ when the bulk velocity ratio $V_L/V_T$ is equal to $2n/(2m+1)$, and for anti-symmetrical modes $A_{2m+1}$ and $A_{2n}$ when $V_L/V_T=(2m+1)/2n$. Except for these particular cases, in the vicinity of a cut-off frequency $(kd \ll 1)$, the dispersion law $\omega(k)$ can be developed to the second order as 
\begin{equation}
	\omega(k) = \omega_c + D k^2 + O(k^3).
\end{equation}
For a thickness-shear resonance, $(k=0)$, the normal surface displacement and the conversion coefficient vanish: $C_{th}(0)=0$. Then, the first order Taylor expansion of this coefficient can be written as 
\begin{equation}
	C_{th}(k)=C k + O(k^2).
	\label{Eq.Cth}
\end{equation}
The normal surface displacement can be approximated as 
\begin{equation*}
	u(r,t)=\frac{1}{2 \pi} C Q(\omega_c)e^{i\omega_c t} \int_0^{+\infty}{ B(k)e^{iD k^2 t} k^2 \, dk},
\end{equation*}
or 
\begin{equation}
	u(r,t) = \frac{1}{2 \pi} C Q(\omega_c) e^{i\omega_c t} B(0) \int_0^{+\infty}{e^{(iD t-s) k^2} k^2 \, dk}.
\end{equation}
Using the parameter $u=k^2$, and the Laplace transform~\cite{Ryzhik07} equation $\int_0^{+\infty} {e^{-xu} u^{1/2}\,du} = \Gamma(3/2)x^{-3/2}$ with $\Re (x)>0$ for $x=s-iDt$, the displacement can be approximated by 
\begin{equation*}
	u(r,t) = \frac{\Gamma(3/2)}{4 \pi} C Q(\omega_c) B(0)\frac{e^{i\omega_c t}} {(s-iDt)^{3/2}}
\end{equation*}
For times $ t\gg s/D$ the displacement is
\begin{equation}
	u(r,t) = \frac{\Gamma(3/2)}{4\pi (Dt)^{3/2}} C Q(\omega_c) B(0) e^{i\left(\omega_c t + \frac{3\pi}{4}\right)} \sim (Dt)^{-1.5}
	\label{sheardecay}
\end{equation}
It results that the decay of thickness mode resonance is much faster that the decay of the ZGV mode resonance, this point will be discussed in the next section. In order to explain why the $A_3$ resonance is detected while, for example, the $A_1$ or the $A_5$ are not, it is necessary to compare the relative amplitude of the resonances and thus to estimate the coefficients $D$ and $C$. The dependence of $D$ and $C$ with respect to the Poisson's ratio and the order of the resonance are now investigated. 
 
% --------------------------------------------------------------------------------
\subsection{Curvature of the dispersion law at thickness resonance}\label{sec:13}
% --------------------------------------------------------------------------------

\begin{figure}[!ht]
\centering
\subfigure{\includegraphics[width=\columnwidth]{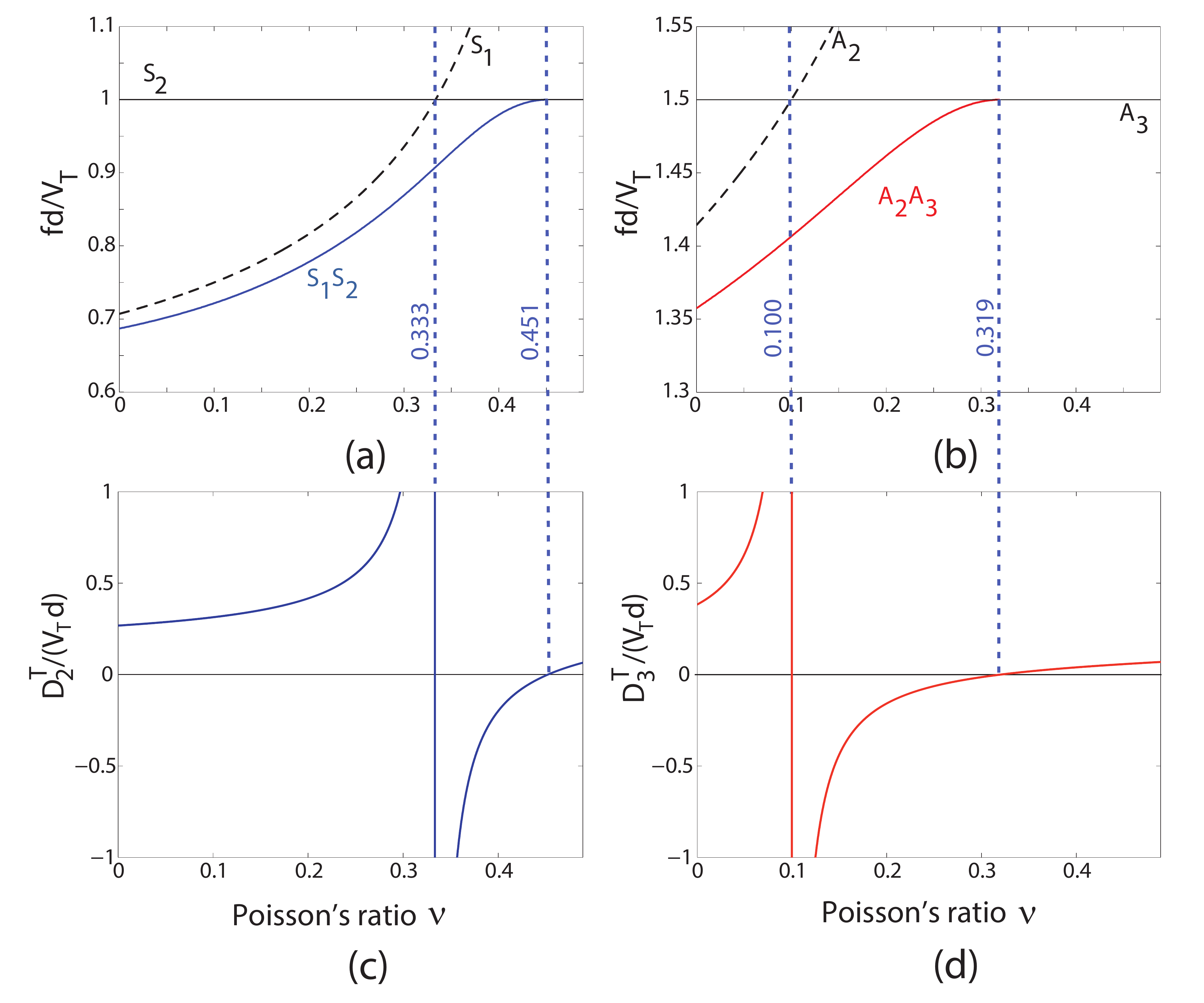}\label{curvature_shear1}}
\subfigure{\label{curvature_shear2}}
\subfigure{\label{curvature_shear3}}
\subfigure{\label{curvature_shear4}}
\caption{Normalized resonance frequencies for $S_1S_2$-ZGV modes and $S_1$ stretch and $S_2$ thickness-shear modes (a) and for $A_2A_3$-ZGV modes and $A_3$ shear and $A_2$ thickness-stretch modes (b). Normalized curvature of the thickness-shear modes $S_2$ (c) and $A_3$ (d) as function of the Poisson's ratio. The vertical lines indicate coincidence frequencies and upper limits of ZGV modes.}
\label{curvature_shear}
\end{figure}

For thickness resonances, the curvatures of the dispersion curve were derived by Mindlin and can be found in Shuvalov and Poncelet.\cite{Shuvalov08} The coefficient $D^T_n$ of a thickness-shear resonance at frequency $f_c=nV_T/2d$ is 
\begin{equation}
	D^T_n = \frac{V_T d}{2\pi n} \left(1+\frac{16}{n\pi} \left(\frac{V_T}{V_L}\right) \tan\left[{\frac{n\pi}{2}\left(1-\frac{V_T}{V_L}\right)}\right]\right)
	\label{curvatureDnT}
\end{equation}  
The normalized curvature $\delta(\nu)=D/V_T d$ of the dispersion curve at the origin $k=0$ for the $S_2$ and the $A_3$ thickness-shear resonances are displayed in Fig.~\ref{curvature_shear}.\\

For the $A_3$ mode, the curvature diverges for $\nu= 0.1$ ($V_T/V_L=2/3$) when the $A_3$ shear resonance intersects the $A_2$ stretch resonance [Fig.~\ref{curvature_shear2}]. Above this point, the curvature is negative [Fig.~\ref{curvature_shear4}], which means there is a backward wave branch emanating from the thickness-shear mode and ending on the ZGV mode. Then, the curvature vanishes for $\nu=0.319$, which corresponds to the upper limit of the ZGV mode. At this point, according to Eq.~\eqref{ZGVdecay}, the amplitude of the thickness resonance diverges. The Duralumin Poisson's ratio is just above this limit and the curvature is $\delta_{A_3}=0.01$ which leads to $\delta_{A_3}^{-1.5}=1000$. On the contrary, for the $S_2$ mode the curvature is much larger [Fig.~\ref{curvature_shear3}], $|\delta_{S_2}|>1$ which explains why the $S_2$ thickness-shear resonance cannot be detected. For thickness-stretch resonances of order $n$, the curvatures $D_n^L$ found in~\cite{Shuvalov08} are equal to
\begin{equation}
	D^L_n=\frac{V_Ld }{2\pi n} \left(1+\frac{16}{n\pi} \left(\frac{V_T}{V_L}\right)^3 \tan\left[{\frac{n\pi}{2}\left(1-\frac{V_L}{V_T}\right)}\right]\right)
	\label{curvatureDnL}
\end{equation}

% ------------------------------------------------------------
\subsection{Conversion coefficient $C_{th}$}\label{sec:14}
% ------------------------------------------------------------

The dependence of the conversion coefficient $C_{th}$ on the Poisson's ratio is complex. Here, we present a simplified approach for thickness modes. For a metallic surface and in the thermo-elastic regime, the source mostly induces an in-plane force. Thus, one can consider that the conversion coefficient is proportional to the in-plane component $u_P$ of the mode at the surface. As the normal displacement $u_N$ is detected at the surface, the conversion coefficient can be defined as the product of the two components normalized to the integral of the squared displacement through the plate thickness 
\begin{equation}
	C_{th}(k) = \frac{ |u_P(h,k)|\,|u_N(h,k)|}{||\mathbf{u}||^2 },
\end{equation}
where $h$ is the half-plate thickness $(-h \leqslant x_2 \leqslant h)$. In the vicinity of a thickness-shear mode we have $|u_P(x_2,k)|\gg |u_N(x_2,k)|$ and
\begin{equation*}
   \begin{array}{ll}
	||\mathbf{u}||^2 &= \int_{-h}^h \left(|u_P(x_2,k)|^2 + |u_N(x_2,k)|^2 \right) dx_2 \\
	&\approx	|u_P(h,k)|^2 \int_{-h}^h sin^2\left(l\frac{\pi}{2}\frac{x_2}{h}\right) dx_2 = h |u_P(h,k)|^2
   \end{array}
\end{equation*}
The conversion coefficient simplifies into
\begin{equation*}
	C_{th}(k) \approx  \frac{|u_N(h,k)|}{ h|u_P(h,k)|}
\end{equation*}
The general expressions of the in-plane $(u_P)$ and normal $(u_N)$ displacements of symmetrical $(\alpha=0)$ and anti-symmetrical $(\alpha=\pi/2)$ modes are given by~\cite{Royer99a}
\begin{equation}
\small{
  \left\{
    \begin{array}{ll}
		u_P(x_2,k) &= -ikB\cos{(px_2+\alpha)} +qA\cos{(qx_2+\alpha)}\\
		u_N (x_2,k) &= -pB\sin{(px_2+\alpha)}+iAk\sin{(qx_2+\alpha)}
	\end{array}
  \right.
  }
  \label{displacements}
\end{equation}
where $p^2=(\omega/V_L)^2-k^2$ and $q^2=(\omega/V_T)^2-k^2$. The coefficients $A$ and $B$ depend on $k$ and satisfy the two equations
\begin{subequations}
    \small{
 	\begin{empheq}[left=\empheqlbrace]{align}
  		(k^2-q^2)B\cos(ph+\alpha)+2ikqA\cos(qh+\alpha) &= 0  \label{ConditionLimites:a}\\
    	2ikpB\sin(ph+\alpha)+(k^2-q^2)A\sin(qh+\alpha) &= 0  \label{ConditionLimites:b}
	\end{empheq}}
\end{subequations}
In the vicinity of the thickness-shear resonance of order $l$ we have the following Taylor expansion 
\begin{equation*}
	\begin{array}{ll}
  		qh = \frac{\omega_c h}{V_T} +O(k^2) &= l\frac{\pi}{2} +O(k^2)\\
  		ph = \frac{\omega_c h}{V_L} +O(k^2) &= l\frac{\pi}{2}\frac{V_T}{V_L} +O(k^2)
	\end{array}
\end{equation*}
For a symmetrical ($\alpha=0$) shear resonance of order $l=2n$, we have $\sin(qh+\alpha)=O(k^2)$ and $\cos(qh+\alpha)=(-1)^n+O(k^2)$ and Eq.~\eqref{ConditionLimites:a} gives $l\pi B\cos[l(\pi/2)(V_T/V_L)+\alpha] = 4ikh(-1)^n A$. The expressions of the displacements at the surface become
\begin{equation*}
\left\{
  \begin{array}{ll}
	|u_P(h,k)|  &=  A\, l\frac{\pi}{2h}  + O(k^2)\\
	|u_N (h,k)| &=  2Ak \frac{V_T}{V_L}\left| \tan{\left(l\frac{\pi}{2} \frac{V_T}{V_L}+ \alpha\right)}\right| + O(k^2)
  \end{array}
\right.
\end{equation*}
For an anti-symmetrical shear resonance ($\alpha=\pi/2$) of order $l=2m+1$, analogous equations can be written. Finally, the coefficients of the leading term of $C_{th}$ [Eq.~\eqref{Eq.Cth}] is
\begin{equation}
	C^T_{l} = \frac{4}{l\pi}  \frac{V_T}{V_L} \left|\, \tan\left(l\frac{\pi}{2}\frac{V_T}{V_L}+\alpha\right)\right|
\end{equation}

A similar derivation for the thickness-stretch modes is given in appendix~\ref{sec:A1} and leads to the following coefficients
\begin{equation}
	C^L_l =  \frac{4}{l\pi}  \frac{V_T}{V_L} \, \left|\cot\left(l\frac{\pi}{2}\frac{V_L}{V_T}+\alpha\right)\right|
\end{equation}

\begin{figure}[!ht]
\centering
\subfigure{\includegraphics[width=0.88\columnwidth]{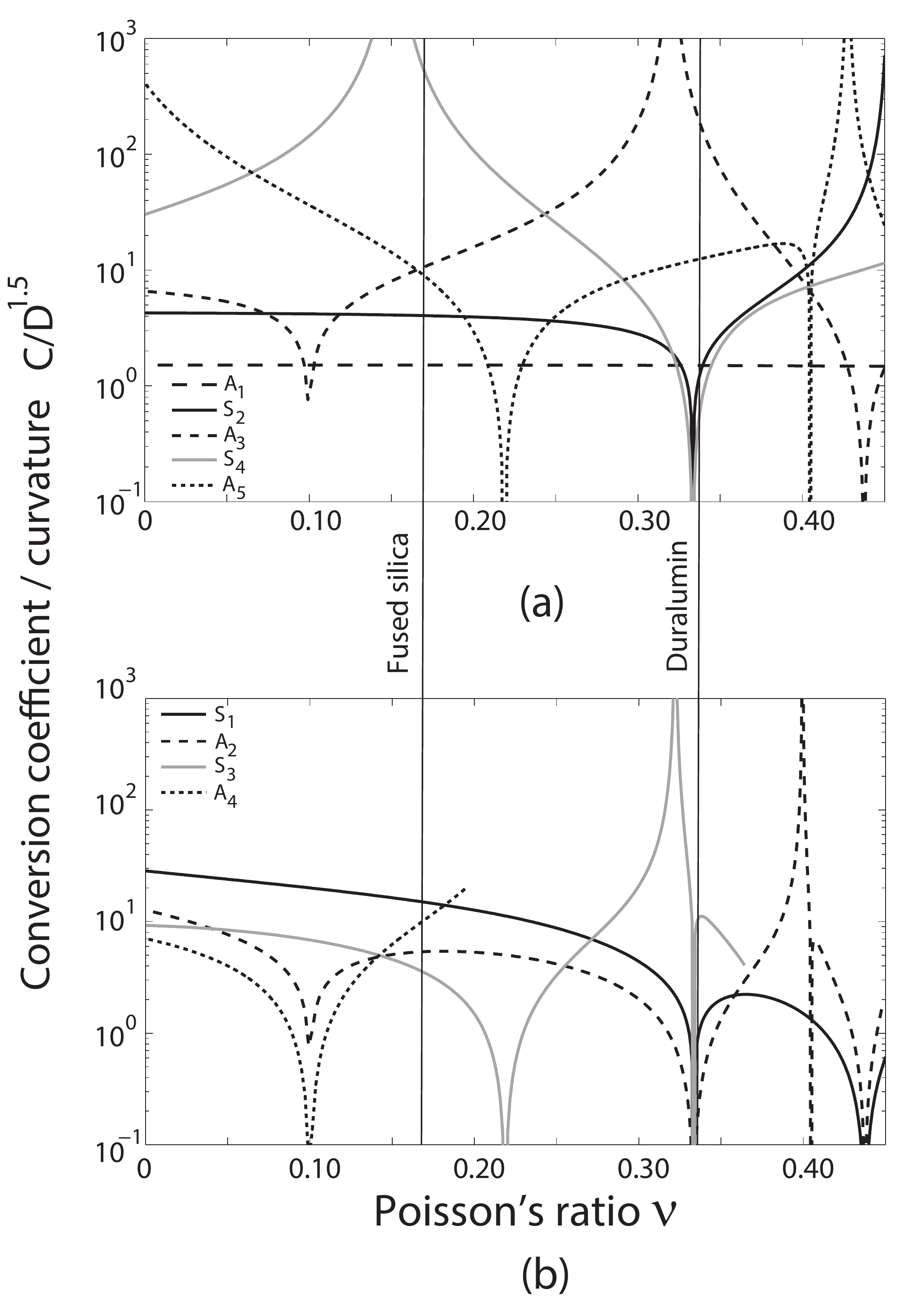}\label{CLCT1}}
\subfigure{\label{CLCT2}}
\caption{(a) The constant $C/D^{1.5}$ as a function of the Poisson's ratio for the first five thickness-shear resonances ($A_1$, $S_2$, $A_3$, $S_4$, $A_5$). (b) The constant $C/D^{1.5}$ as a function of the Poisson's ratio for the first four thickness-stretch resonances ($S_1$, $A_2$, $S_3$, $A_4$) of frequencies below $5V_T/2d$.}
\label{CLCT}
\end{figure}

These coefficients diverge if $V_T/V_L$ is rational which corresponds to a coincidence between thickness-shear and thickness-stretch resonances. As mentioned in Sec. \ref{sec:12}, the model is no more valid in this case. According to Eq.~\eqref{ZGVdecay}, the amplitude of the resonance is proportional to $C/D^{1.5}$. This parameter is plotted as a function of the Poisson's ratio in Fig.~\ref{CLCT1} for the thickness-shear modes and in Fig.~\ref{CLCT2} for the first four thickness-stretch modes.\\

\begin{figure*}[!ht]
%\begin{sidewaysfigure}
\centering
\subfigure{\includegraphics[width=\textwidth]{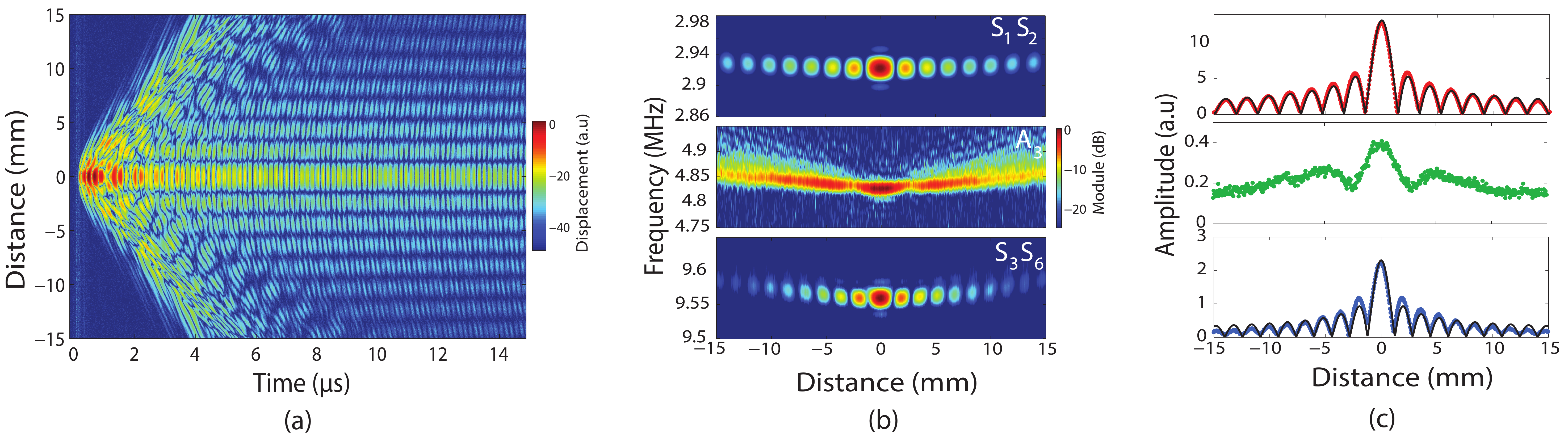}\label{dep_field_profil:a}}
\subfigure{\label{dep_field_profil:b}}
\subfigure{\label{dep_field_profil:c}}
\caption{Normal displacements measured as function of the distance from the source on a 1-mm thick Duralumin plate(colorbar: absolute displacement in logarithmic scale)(a). Temporal Fourier transforms around $S_1S_2$, $A_3$ and $S_3S_6$ resonances frequencies (b). Maximum amplitude (dots) and fit with zero order Bessel function (solid line) (c).}
\label{dep_field_profil}
%\end{sidewaysfigure}
\end{figure*}

For sake of clarity the coefficient where plotted only for resonance frequencies below the fifth shear resonance frequency $f_c < 5V_T/2d$. The two green dash lines indicate the Poisson's ratios of fused silica and Duralumin. It appears that for fused silica the coefficient of the $S_4$ thickness-shear resonance is higher by at least one order of magnitude than for the other modes. It means that the $S_4$ resonance will dominate the other thickness resonances. The same observation can be done for the $A_3$ thickness-shear mode in Duralumin. This graph also indicates that the thickness-stretch resonances should not be easily detected. Is it important to underline that these curves give the relative weights of thickness resonances for a given Poisson's ratio. To compare resonance amplitudes for two different materials, one should  consider other parameters like optical absorption or thermal dilatation also involved in the conversion process.

% ============================================
%    Setup + experimental/numerical results
% ============================================
\section{Experimental results}\label{sec:2}

A heterodyne interferometer~\cite{Royer86} associated with a Q-switched Nd:YAG laser at $\Lambda=1064$~nm was used. Lamb waves were generated in the thermo-elastic regime by the absorption of a short pulse (duration $\Delta = 20$~ns, energy $3.1$~mJ, repetition rate of $20$~Hz, spot diameter 2.5 mm). The calibration factor for mechanical displacements normal to the surface ($10$~mV/nm) is constant over the detection bandwidth ($20$~kHz to $45$~MHz). Measurements were made on two materials, Duralumin and fused silica, having a low attenuation (less than 1~dB/m in the MHz frequency range). For fused silica, a thin aluminium layer was deposited for a sufficient optical absorption at the surface.

% ----------------------------------------------------------------------------------------------
\subsection{Displacement profile at ZGV and thickness-shear resonance frequencies}\label{sec:21}
% ----------------------------------------------------------------------------------------------

The spatial profile of the resonances was measured on a 1-mm thick Duralumin plate. The ratio $R/d$ is equal to 1.25 which corresponds to $R \sim 0.32\lambda_{S_1S_2}$. The normal surface displacements, detected by scanning the probe across the source along a 30-mm line, is displayed in Fig.~\ref{dep_field_profil:a} for the first $15~\mu s$. After about $8~\mu s$ most propagating modes have escaped the scanned area. A standing mode is clearly observed immediately after. The temporal Fourier transforms of these signals calculated on the same time-window ($7$ to $80\,\mu s$) reveal 3 resonances at 2.923~MHz, 4.827~MHz and 9.560~MHz. As expected, they correspond respectively to $S_1S_2$, $A_3$ and $S_3S_6$ resonances. The amplitude distributions around the resonance frequencies are displayed in Fig~\ref{dep_field_profil:b}. The type of resonance can be distinguished from the spatial profiles.\\

For ZGV resonances, they follow a Bessel function $J_0(2\pi r/\lambda)$ with $\lambda=3.85$~mm for $S_1S_2$ and $\lambda=3.20$~mm for $S_3S_6$, corresponding to the interference of backward and forward waves of opposite phase velocities. The profile of the thickness resonance is different and only corresponds to propagative forward modes. Contrary to ZGV modes, the profile depends on the time window used in the Fourier transform especially at short times.

% ------------------------------------------------------
\subsection{Resonance temporal decays}\label{sec:22}
% ------------------------------------------------------

\begin{figure}[!ht]
\centering
\subfigure{\includegraphics[width=\columnwidth]{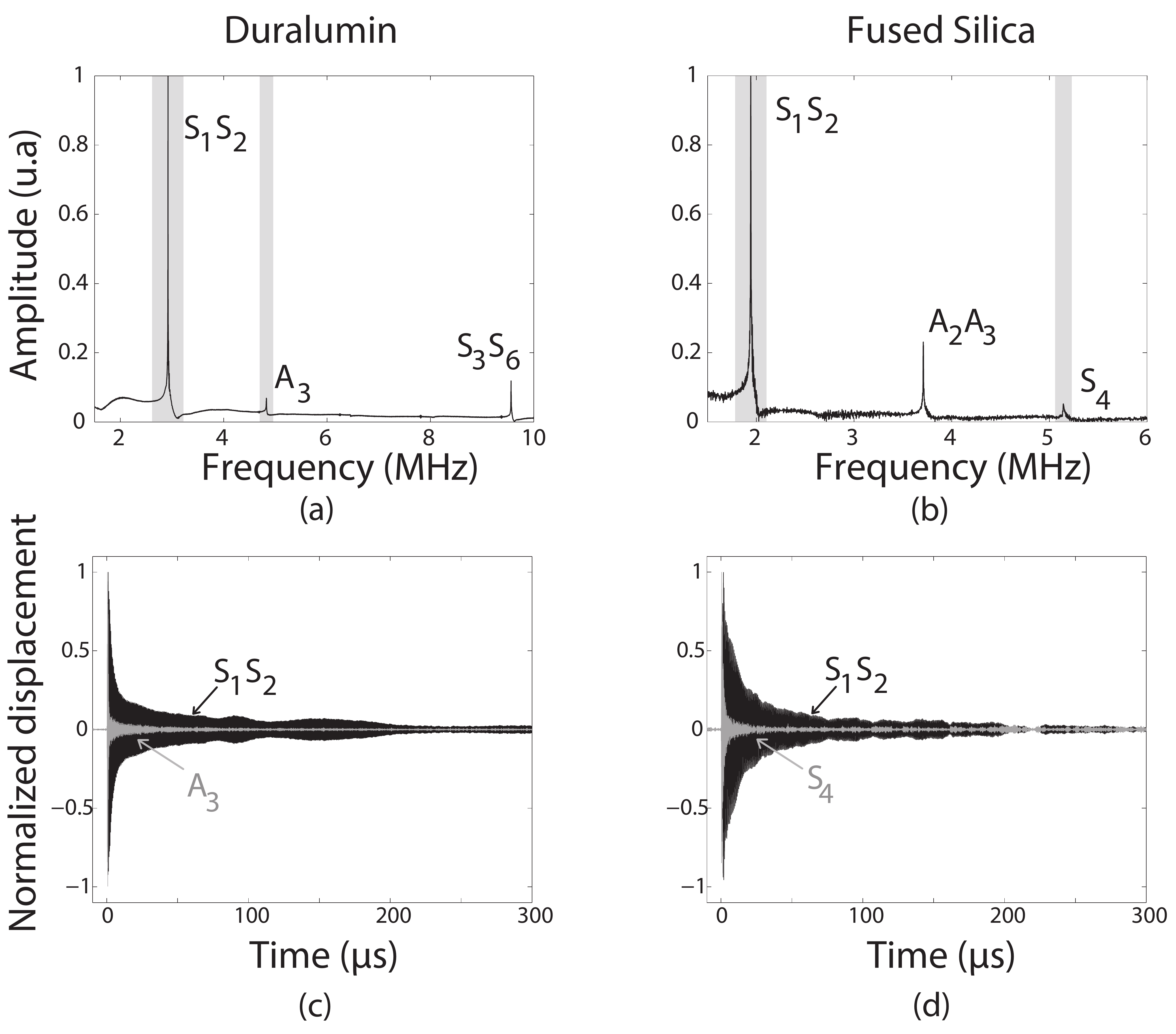}\label{att_zgv_modes1}}
\subfigure{\label{att_zgv_modes2}}
\subfigure{\label{att_zgv_modes3}}
\subfigure{\label{att_zgv_modes4}}
\caption{Normal surface displacement generated and detected at the same point on a 1-mm thick Duralumin (left) and fused silica (right) plates: (a) and (b) Fourier transforms, (c) and (d) signals filtered around $S_1S_2$-ZGV frequencies (black line) and around the $A_3$ thickness frequency for Duralumin and around $S_4$ thickness frequency for fused silica (grey line). }
\end{figure}

The normal surface displacements were detected at the source location during $300~\mu s$ on both Duralumin and fused silica plates. The Fourier transforms of these signals are displayed in Figs.~\ref{att_zgv_modes1} and \ref{att_zgv_modes2}. In both cases, the $S_1S_2$ mode dominates. According to the previous analysis, higher order resonances for fused silica are different than for Duralumin and correspond to $A_2A_3$-ZGV mode and $S_4$ thickness-shear mode. The signals filtered (bandwidth $0.3$~MHz) around the resonance frequencies of the $S_1S_2$-ZGV mode and thickness-shear modes $A_3$ for Duralumin and $S_4$ for fused silica are displayed in Figs.~\ref{att_zgv_modes3} and \ref{att_zgv_modes4}.\\

\begin{figure}[!ht]
\centering
\subfigure{\includegraphics[width=\columnwidth]{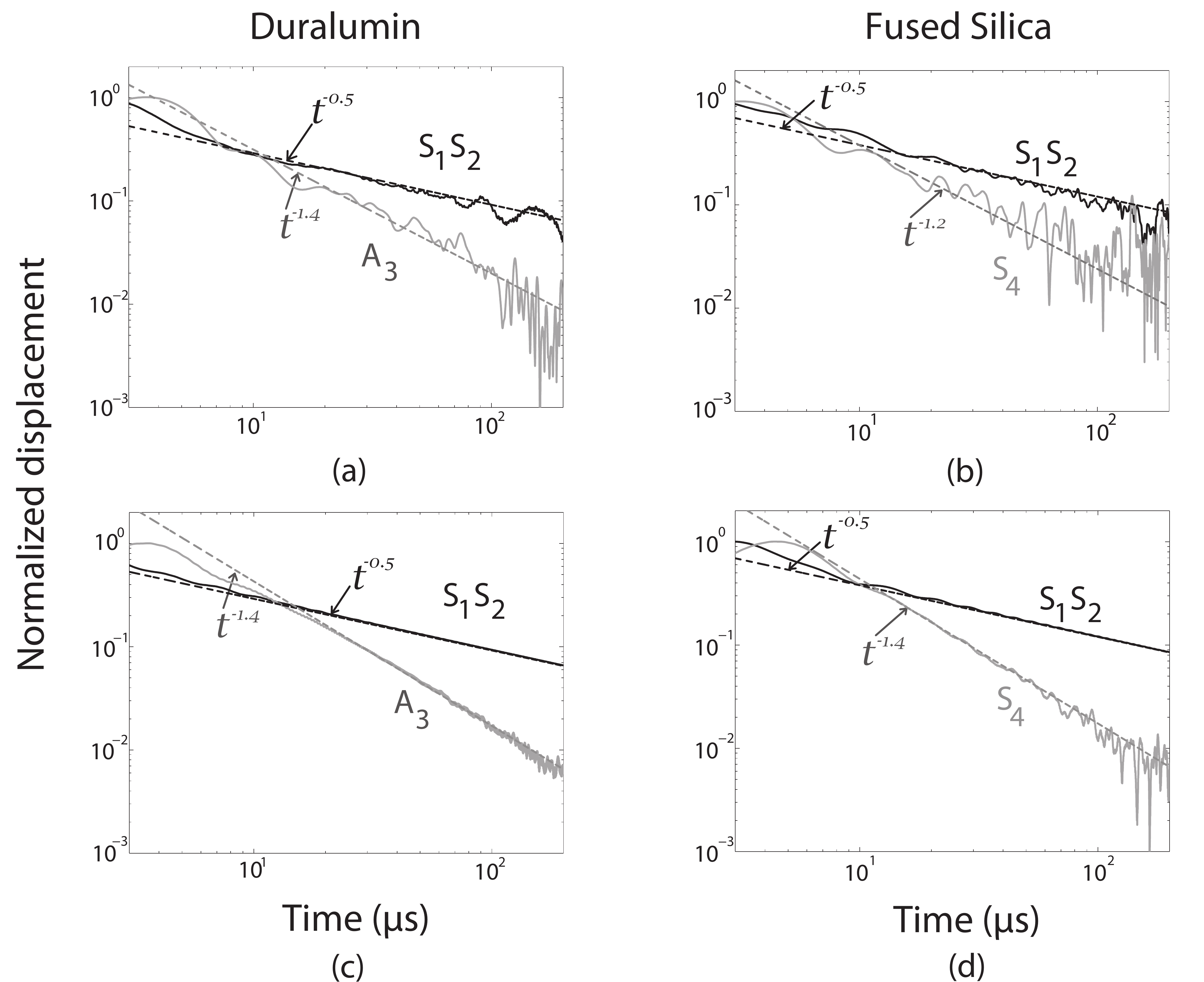}\label{att_zgv_log1}}
\subfigure{\label{att_zgv_log2}}
\subfigure{\label{att_zgv_log3}}
\subfigure{\label{att_zgv_log4}}
\caption{ Experimental (a, b) and numerical (c, d) temporal power law decays in logarithmic scales.}
\end{figure}

It is clearly observed that amplitudes of thickness modes decay much faster than ZGV mode amplitudes. The temporal decays were estimated by using a Hilbert transform of the filtered signals. As shown in Figs.~\ref{att_zgv_log1} and~\ref{att_zgv_log2}, the amplitudes decrease like $t^{-0.5}$ for $S_1S_2$-ZGV modes. Similar decays were observed for $S_3S_6$-ZGV mode in Duralumin and $A_2A_3$-ZGV mode in fused silica. This result is in good agreement with the theory [Eq.~\eqref{ZGVdecay}]. For thickness-shear modes, the power law fit provides a $-1.4$ exponent for Duralumin and $-1.2$ for fused silica, which is slightly different than the $-1.5$ predicted by the above theory [Eq.~\eqref{sheardecay}]. Using a homemade finite difference code, the local response was also calculated and similar temporal decays ($-1.4$) were obtained for both resonance types as shown in Figs.~\ref{att_zgv_log3} and~\ref{att_zgv_log4}. This decay is observed for time higher that $10~\mu s$, which is consistent with the validity condition $t\ll sD$, since $sD$ is equal to $2~\mu s$ for the $A_3$ mode in the Duralumin plate. For thickness modes, the discrepancy between experiments and theory could be ascribed to the various approximations made in the model.\\

% ==============
%   Conclusion
% ==============
\section{Conclusion}

The local elastic resonances of a plate excited by a laser source was studied for low attenuation materials. It was observed that while the dominant resonances are associated to zero group velocity Lamb mode of finite wavelength, some thickness resonances are also detected. The type of resonance can be determine from the spatial profiles. For ZGV resonances, it follows a Bessel function corresponding to the interference of counter propagating modes of opposite phase velocities. The profile of thickness resonance is very different and only correspond to interferences of forward modes. In the experiment, the $A_3$ thickness-shear resonance is detected in Duralumin plates and the $S_4$ thickness-shear resonance is detected in fused silica, while the other thickness modes are not observed. This is explained through a simplified model which shows that the amplitude of the resonances is proportional to $D^{-1.5}$ where $D$ is the curvature of the dispersion law and that the temporal decay of the resonance is also different: for ZGV modes it follows a power law $t^{-0.5}$, while for thickness resonances the decay is much faster. The proposed theory predicts that the decay is as $t^{-1.5}$ while experimental and numerical results reveal slightly slower decay as $t^{-1.4}$. This discrepancy can be ascribed to the approximations made in the theory. The development of a more complete model should be the object of further studies.\\

%===========
 \appendix
%===========

\section{Coefficient of thermo-elastic conversion for the thickness-stretch modes}\label{sec:A1}

Following the same approach than for thickness-shear resonances, the conversion coefficient $C_{th}$ for thickness-stretch resonances can be approximated as
\begin{equation*}
C_{th}(k) \approx \frac{ |u_P(h,k)|}{h|u_N(h,k)|}
\end{equation*}
For a resonance of order $l$: $\omega_ch = l V_L\pi$, and the following Taylor expansions can be written
\begin{equation*}
	\left\{
		\begin{array}{ll}
			qh = \frac{\omega_c h}{V_T} + O(k^2) &= l\frac{\pi}{2} \frac{V_L}{V_T} + O(k^2)\\
			ph = \frac{\omega_c h}{V_L} + O(k^2) &= l\frac{\pi}{2} + O(k^2)
		\end{array}
	\right.
\end{equation*}
For a symmetrical resonance of order $l=2m+1$ and $\alpha=0$ , we have $\sin(ph+\alpha)= (-1)^m + O(k^2)$ and $\cos(ph+\alpha) = O(k^2)$. Using the first order development and inserting the above equation in Eq.~\eqref{ConditionLimites:b}, it comes
%$$2ik(-1)^m  pB = q^2Asin(qh+\alpha)$$
%
\begin{equation*}
	2ik(-1)^m  B =  l\frac{\pi}{2} \left(\frac{V_L}{V_T}\right)^2 A\sin\left(l\frac{\pi}{2} \frac{V_L}{V_T}+\alpha\right),
\end{equation*}
and the parallel and normal displacements become
%
%\begin{widetext}
\begin{equation*}
	\left\{
		\begin{array}{ll}
			u_P(h,k)&=\,qA\cos(qh+\alpha) \\
			        &= -(-1)^m 2ikB\,\frac{V_T}{V_L}\,\cot{\left((2m+1)\frac{\pi}{2} \frac{V_L}{V_T}+\alpha\right)}\\
			u_N(h,k)&= -pB\sin(ph+\alpha) \\
			        &= \,(-1)^m B\,(2m+1)\frac{\pi}{2h}
		\end{array}
	\right.
\end{equation*}
%\end{widetext}
%
It follows that the coefficient $C^L_{2m+1}$ is 
\begin{equation}
	C^L_{2m+1} =  \frac{4}{(2m+1)\pi } \frac{V_T}{V_L} \, \cot\left((2m+1)\frac{\pi}{2}\frac{V_L}{V_T}\right)
\end{equation}
For an anti-symmetrical resonance of order $l=2n$ and $\alpha=\pi/2$, we have $\sin(ph+\alpha)= -(-1)^n+O(k^2)$ and $\cos(ph+\alpha)=O(k^2)$. Using the first order development of Eq.~\eqref{ConditionLimites:b} we have 
\begin{equation*}
	2ik(-1)^n  B =  l\frac{\pi}{2} \left(\frac{V_L}{V_T}\right)^2 A\sin\left(l\frac{\pi}{2} \frac{V_L}{V_T}+\alpha\right)
\end{equation*}
The displacements become
\begin{equation*}
	\begin{array}{ll}
	u_P(h,k) &= \,(-1)^n 2ikB\,\frac{V_T}{V_L}\,\cot\left(\frac{\pi}{2} \frac{V_L}{V_T}+\alpha\right)\\
	u_N (h,k) &= -(-1)^n B\,l\frac{\pi}{2h}\\
	\end{array}
\end{equation*}
and the coefficient of the leading term of $C_{th}$
\begin{equation}
	C^L_{2n} = \frac{4}{2n\pi}\frac{V_T}{V_L} \,\cot{\left(2n\frac{\pi}{2} \frac{V_L}{V_T}+\alpha\right)}
\end{equation}
More generally, we obtain for thickness-stretch modes of order $l$:
\begin{equation}
	C^L_l = \frac{4}{l\pi}\frac{V_T}{V_L} \,\cot\left(l\frac{\pi}{2} \frac{V_L}{V_T}+\alpha\right)
\end{equation}
 \\
% ============================
\section*{Acknowledgements}
% ============================

This work was supported by LABEX WIFI (Laboratory of Excellence ANR-10-LABX-24) within the French Program ``Investments for the Future'' under reference ANR-10-IDEX-0001-02 PSL*.

% =================
%   Références
%  \section*{References}
% =================

%\bibliographystyle{custom}         % Titre complet
%\bibliographystyle{apsrev}         % or apsrmp  % Pas de titres
%\bibliography{Biblio/BiblioZGV}

\end{document}